*This internet, on the ground*
*Nick Merrill*
*December 9, 2021*

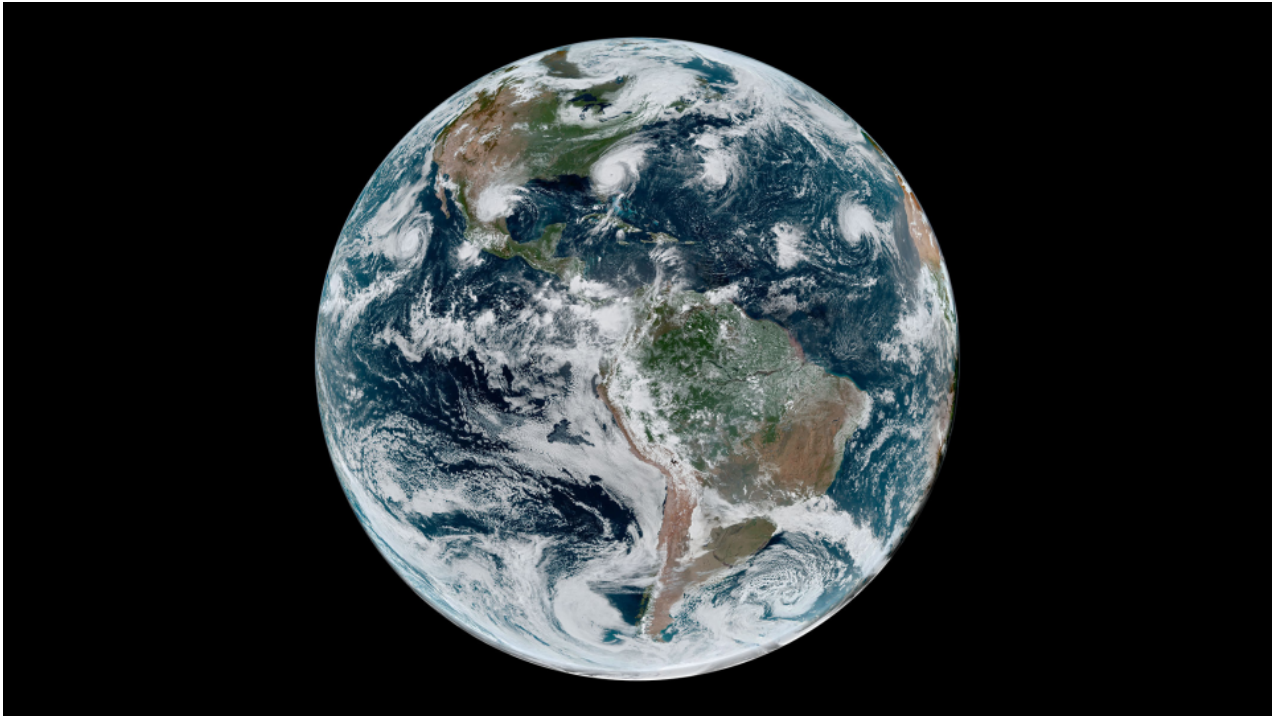

Figure 1: The Internet, as seen from space. *NASA*.

The internet's key points of global control lie in the hands of a few people, primarily private organizations based in the United States. These control points, as they exist today, raise structural risks to the global internet's long-term stability. I argue: the problem isn't that these control points exist, it's that there is no popular governance over them. I advocate for a *localist* approach to internet governance: small internets deployed on municipal scales, interoperating selectively, carefully, with this internet and one another.

This (Figure 1) is my favorite photo of the internet. It reminds me this internet is finite. Yes, it feels endless when you're browsing it, but there is, indeed, only so much internet. Perhaps this photo is missing a few deep-space probes, but, materially, the internet *does* end. There is an inside, and an outside. I hope that relaxes you.

Now, I called this paper *This internet, on the ground*, and here I am showing you the earth from space. The literal opposite of the ground. But it's good to have perspective, and perspective is ultimately what I want to impart. I want to help you experience this internet as something that *is* on the ground.

To do that, it makes sense to start from space. This view—the planetary view—is how the internet's early proponents saw it. They had transnational aspirations, global aspirations.

And if we don't understand how they saw it, we have no chance of understanding *this* (Figure 2), a grassroots internet in Havana that runs through the streets, peer to peer, grandmas restarting computers to make routing messages propagate.[1] This is a different internet, built in a different way. Adopted to the local material conditions (think: trade embargoes). Structured by a different ideology, an ideology that cared about fundamentally different things.

This internet is global in its reach and impact. This internet is a common, like this planet is a common.

[1] Michaelanne Dye et al. If it Rains, Ask Grandma to Disconnect the Nano: Maintenance and Care in Havana's StreetNet. *ACM Conference on Computer Supported Cooperative Work (CSCW)*, 2019

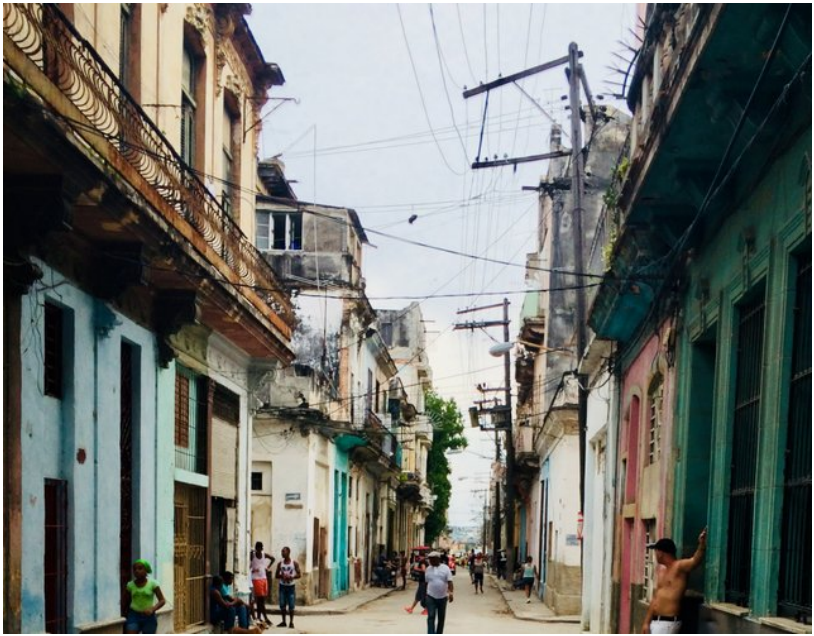

Figure 2: Cables run through the Streets of Havana. From Dye et al.,2019.

What is an internet? In broad strokes, an internet is a mechanism that connects computers. Crucially, this mechanism ought to be scale-free: it should be able to connect any number of computers.

This internet is not the only possible internet. It is only one of many internets that could have been designed,[2] that could have become global.[3] Indeed, other internets predated this one: Mintel in France, Akademset in the U.S.S.R. But this is the big one. This is the hegemonic one. This is *the* internet.

*Why is this one*? Of all the possible internets, why is *this* the one you're using? Once we understand that, we can understand a lot better what the problems facing this internet are, where they come from, and how we can go about solving them.

*An* internet is the *idea* of a scale-free computer network. *This* internet—the one that delivered this document to you—is a specific implementation of this idea.

[2] David D Clark. *Designing an Internet*. MIT Press, 2018
[3] Paul Dourish. Not the internet, but this internet: how othernets illuminate our feudal internet. In *Proceedings of The Fifth Decennial Aarhus Conference on Critical Alternatives*, pages 157–168, 2015

*Why this internet?*

> The LSD we took as a tonic of psychic liberation turned out to have been developed by CIA researchers as a weapon of the Cold War. We had gone to a party in La Honda in 1963 that followed us out the door and into the street and filled the world with funny colors. But the prank was on us.
>
> —ROBERT STONE, *Prime Green*

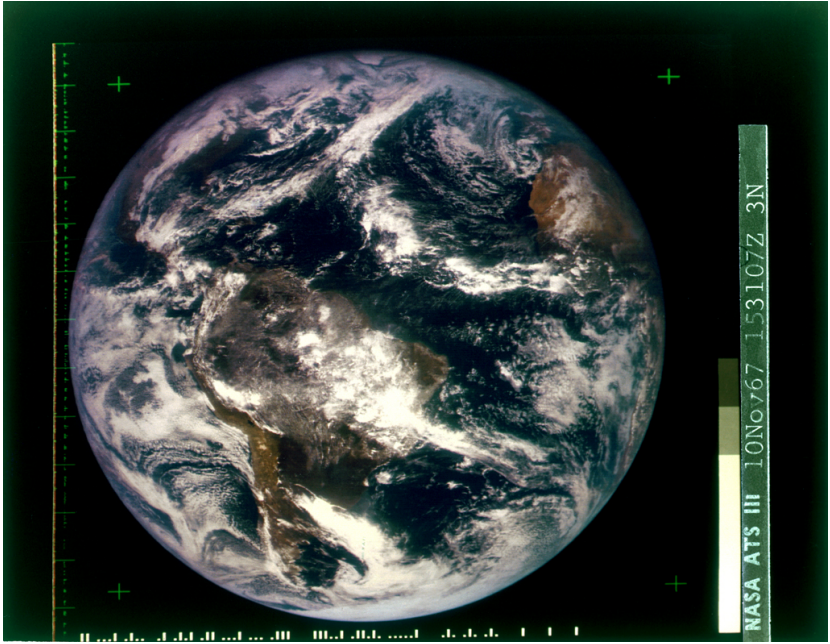

Figure 3: The earth in 1967; a composite of photos taken by the ATS-3 satalite. *NASA.*

To BEGIN, we have to go back to when the planet looked like this (Figure 3). The year is 1967. And this photo—taking this photo—was central to an ongoing power conflict between the United States and the U.S.S.R.[4], one that plays out technologically, primarily.

That's where this internet begins.

[4] Or at least typified by the leadership of those two states.

*Military experiment*

This Internet was a U.S. military project. Its goal was *primarily* to serve the US's international interests: military communications that could withstand kinetic conflict. *Only later* did this internet move into the realm of the domestic—into the realm of the commercial, which is *seen as* the commercial in the eyes of the U.S. policymakers in the 1990s who promote its adoption. But I'm getting ahead of myself. Forget all that. It's 1967.

A few years prior, A RAND corporation employee named Paul Baran had written a paper about communication networks.[5] He was funded by the Air Force, which was interested in building a communication network that *routes around failure*. Something gets bombed, you can still deliver a message.

[5] Paul Baran. On distributed communications networks. *IEEE transactions on Communications Systems*, 12(1):1–9, 1964

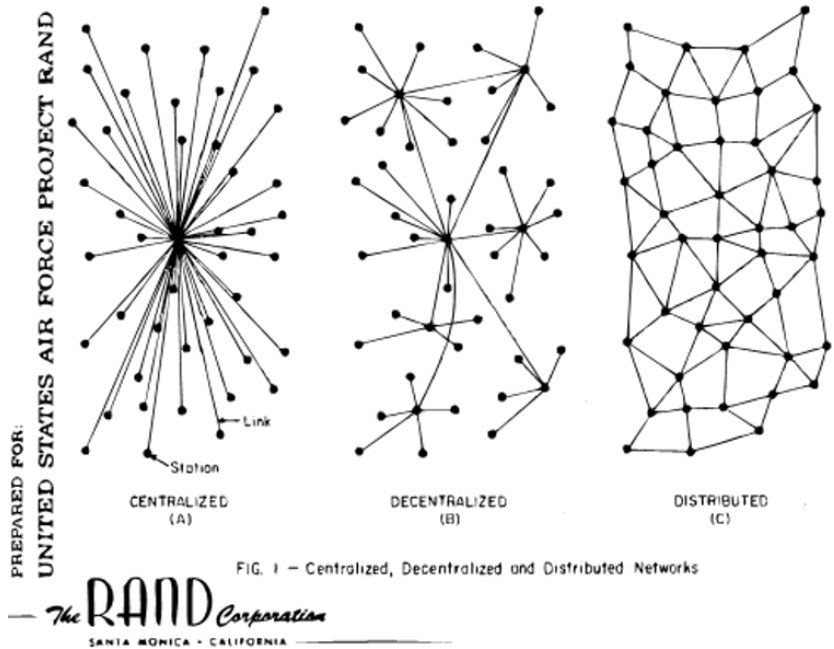

Figure 4: Baran (1964)'s rendition of different structures for communication networks.

He had drawn these three diagrams (Figure 4). A is no good, he argued. If you bomb the center point, no one can deliver a message to anyone. Now, C is the most resilient to attacks. But B splits the difference. It is neither centralized nor distributed: it is *de*centralized; it has no *one* center. Decentralized networks, Baran had argued, are pretty resilient to attacks, while also being a bit more performant, because you can optimize the connector nodes, relaxing the need for every node to relay messages. And he had some graph theory to demonstrate this, some math.

Some people believe the internet is so decentralized as to be politically uncontrollable. To paraphrase Hu (2016), they mistake Baran's diagrams for historical narrative. While this internet has no *single* point of control, it has a finite number of critical ones, many of which—as we'll see in a moment—lie in the United States' legal jurisdiction.

But the U.S. military apparatus *got* (B) in a way that's not mathematical at all: it reminded them of the U.S. highway system, which was considered "hard to bomb" by the Eisenhower-era officials who mostly made up the U.S. Advanced Research Projects Agency (ARPA) leadership at that time. You have a lot of highways, a few connections throughout, and that makes it robust. If one highway gets bombed, you can route trucks around the "outage."

While the comparison between this internet and the highway system have long been questionable,[6] its "fit" as a metaphor is immaterial to the internet's early history. Functionally, this metaphor made this project more than a communications network. It made the project ideological: about pride in this other recent infrastructure project, the interstate highway system, that was good for securitization, good for private industry, good for consumers; about decentralization as a foil to the U.S.S.R.'s command economy.

[6] Tung-Hui Hu. Truckstops on the Information Superhighway: Ant Farm, SRI, and the Cloud. *Journal of the New Media Caucus*, 2014

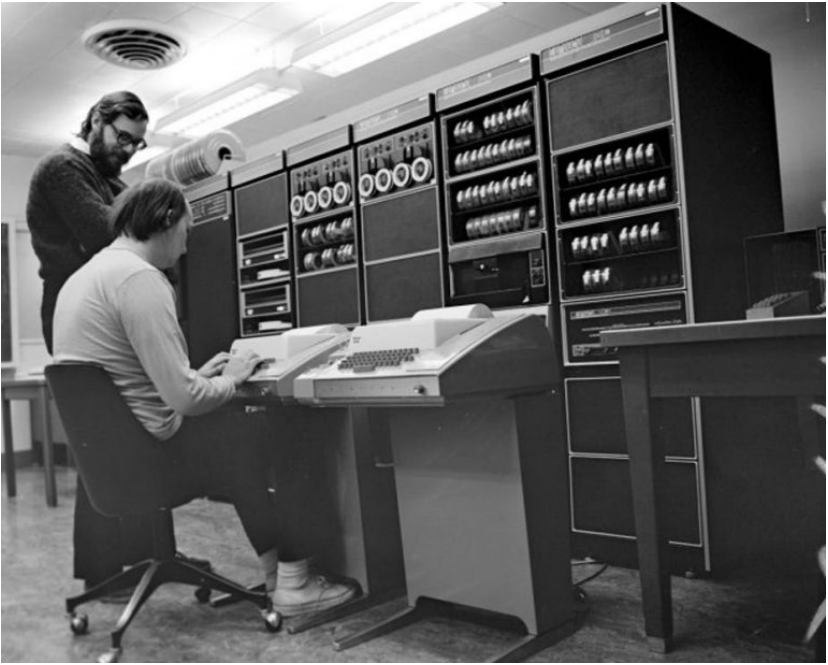

Figure 5: Computing in the 60s. From Giovanni Navarria's How the Internet was born: A stuttered hello. The ARPANET was a *network of networks*. Small local networks are easy to build: we just connect all the computers to a central point (a router). But we can only connect so many computers to a single router. In the ARPANET, a specialized sub-network of computers would be solely responsible for routing traffic between routers (diagram B in Figure 4). An inter-network, an inter-net.

So the idea gained steam and, in 1967, an ARPA research staffer named Larry Roberts draws on Baran's work to build a network of computer researchers [7] He called it the ARPANET.

[7] Cybertelecom. ARPAnet 1966-1968.

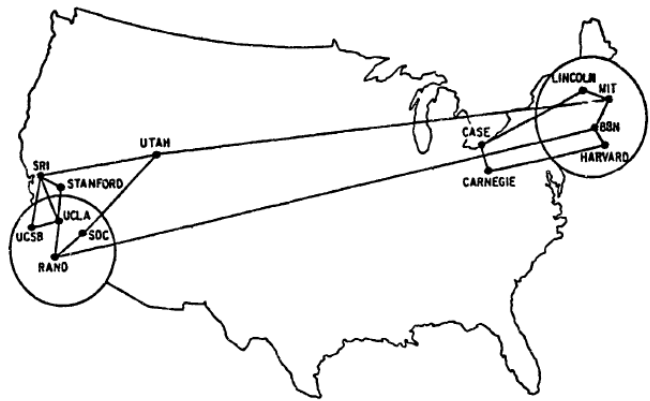

Figure 6: The ARPANET in December 1970.

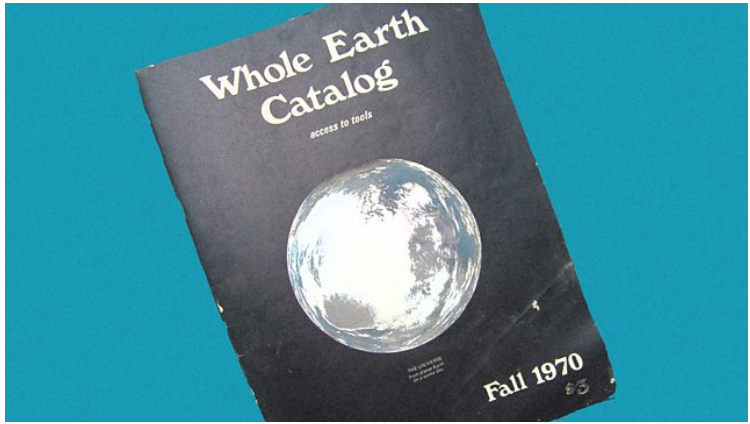

Figure 8: The *Whole Earth Catalog* influenced the internet's early designers who, as Turner (2010) documents, corresponded with Stewart Brand, the publication's editor, to discuss various design decisions. The catalog's cover featured a recently-taken photo of the earth from space (Figure 3).

*Techno-utopians*

The project takes off. It's built in fits and bursts. And, by 1970, the ARPANET runs coast to coast.

But the institutions now administering this network aren't exactly part of the U.S. military establishment (Figure 6). They're research universities, where the ARPANET is maintained mostly by research scientists, university staff. People like me.

For example, people like Jon Postel (Figure 7). They're kind of hippies, but not *such* hippies that they refuse to take money from the U.S. military. And these are the people who started to grow the internet. To scale it up and build it out. They made the thing *work*. They built UNIX commands that are still on your computer. They build the DNS, which is how we access names like "nytimes.com" instead of memorizing some list of IP addresses. In other words, they took this idea of *an* internet and they built *some* internet, some specific internet.

One that—actually kind of worked. It worked well enough that, by 1986, ARPANET grew out of its ARPA funding and became the NSFNet. (NSF is the National Science Foundation, a big, but non-military science funder administered by the U.S. federal government. They tend to fund projects that may have a more domestic interest, or a basic science interest. Ideally a bit of both. This project fits the bill.).

In fact, at this point, it's pretty clear this internet has transnational aspirations.[8] We can make this thing go all the way around the world. Everyone can share this one network, and everyone can address packets to everyone, and this is going to change *everything*—in a way few people understand.[9] But *some* understand: the people who have been maintaining and building out this system since the 1970s.

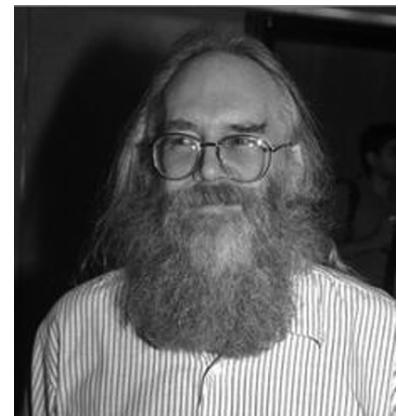

Figure 7: Jon Postel [intentionally rerouted the entire internet in the 1970s](#) to prove a point about the unauthenticated, unilateral nature of internet routing. Today, this basic feature of the Internet's design remains a [serious security flaw](#).

[8] Fred Turner. *From counterculture to cyberculture*. University of Chicago Press, 2010

[9] Pre-internet, the concept of an application that can accept some application-specific logic and send data around through generic packets was not legible to most people. No analogy exists.

These people are deeply complicit in the U.S. government's goals and desires. But, at their core, they're agents of the 1960s counterculture. Many read the Whole Earth Catalog (Figure 8). And This counterculture is, at *its* core, techno-utopian: it believes in the technological to deliver meaningful, lasting social change.

In the 1990s, this attitude calcifies around the idea that a global internet is going to make it *hard to suppress speech*, hard to suppress *free association* between the world's people. The height of 1990s techno-utopianism comes, in my mind, in 1996, when John Perry Barlow gives a speech called *The Declaration of the Independence of Cyberspace*.

> Governments of the Industrial World, you weary giants of flesh and steel, I come from Cyberspace, the new home of Mind. On behalf of the future, I ask you of the past to leave us alone. You are not welcome among us. You have no sovereignty where we gather.
>
> —John Perry Barlow, 1996[10]

Ironically, the venue at which Barlow made this proclamation: Davos, the conference for billionaires. By 1996, national interests are still central to the internet's development, but we already see the capital class congregating around the margins. We'll return to them in a moment.

For now, speech is the thing.

> "The Net interprets censorship as damage and routes around it."
>
> —John Gilmore, 1993[11]

Recall, this is much the same logic the early ARPANET developers used: similar to the US highway system, this network could route around failure. And this perspective, typified by John Gilmore in 1993, percolates through the U.S. Agency for Global Media, which identifies the internet as a helpful tool, particularly for their newly-formed Radio Free Asia arm.[12] These currents, inside and outside of D.C., structure and shape how the Clinton administration envisions the internet.

> "[Beijing] has been trying to crack down on the Internet–good luck. That's sort of like trying to nail Jello to the wall.
>
> –Bill Clinton, 2000 [13]

The historical question at this time was whether authoritarian regimes could survive the internet. (Today, the question is whether the internet can survive authoritarian regimes). This was the "end of history" as it was infamously proclaimed at the time:[14] liberal democracy (such as it is) was sweeping across the globe, and the global spread of the internet *as an idea* is entwined with globalization writ large. Global trade, global commerce, global exchange. And this internet was going to change *all* of that.

---

I like to think
 (it has to be!)
 of a cybernetic ecology
 where we are free of our labors
 and joined back to nature,
 returned to our mammal
 brothers and sisters,
 and all watched over
 by machines of loving grace.

—Richard Brautigan, 1967

I think that I shall never see
A graph more lovely than a tree.
A tree whose crucial property
Is loop-free connectivity.
  —Radia Perlman, 1985

[10] Electronic Frontier Foundation. A Declaration of the Independence of Cyberspace. February 8, 1996.

[11] TIME. First Nation in Cyberspace. 1993

[12] Radio Free Asia. History. Radio Free Asia went on to fund development of Tor, a technology we discuss later.

[13] New York Times. Clinton's Words on China: Trade Is the Smart Thing. March 9, 2000.

[14] Francis Fukuyama. *The end of history and the last man*. Simon and Schuster, 1992

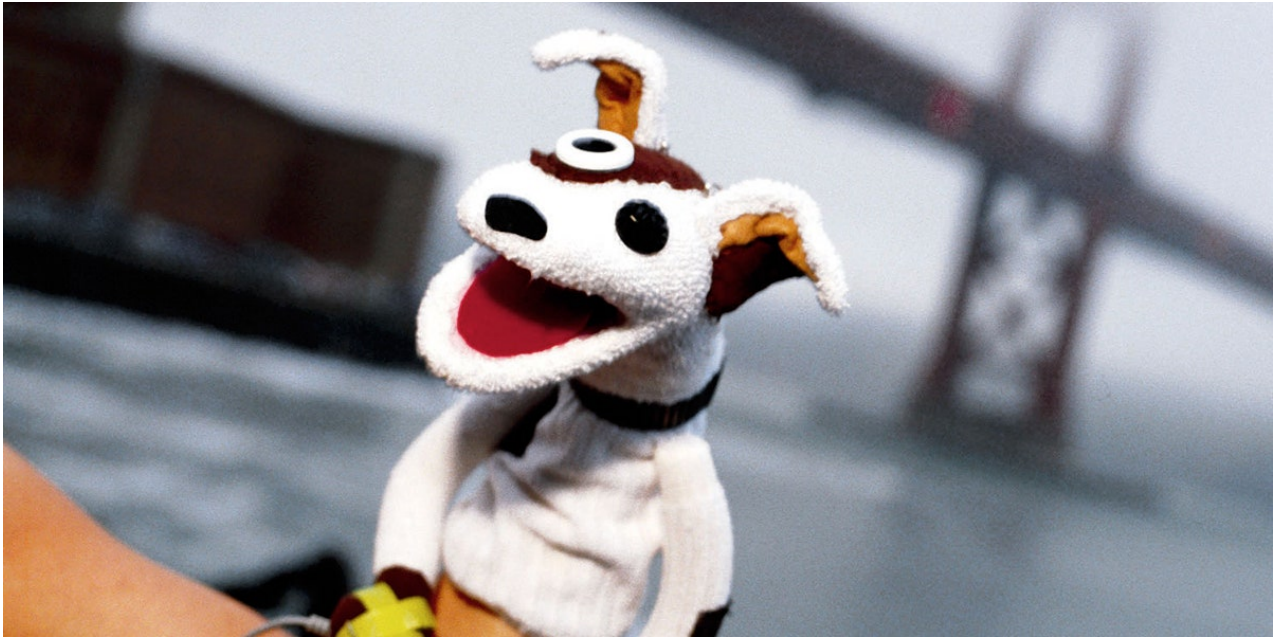

Figure 9: The Pets.com spokespuppet. Pets.com became synonymous with the speculative frenzy of the dot-com boom and bust.

*A feudal internet*

And change it it did.

The "dot-com" boom of the late 1990s was frothy and intense. Senseless valuations. If you're my age, your parents may have lost money in it. But in the long term, the institutions this era shaped have stuck around as cultural forces. Venture capital, the idea of "hockey-stick" growth. This model, and the tech companies it started, have flourished.[15]

In a way—and I win no friends by saying this—those companies built this internet better than the hippies did. Those companies made this internet more stable, more usable, more performant, in many ways more secure than it was—is—by design.

Today, those companies have *become* the internet.[16] Never mind that they own the software infrastructure that provisions internet services, as we'll see in a moment. They increasingly own the physical infrastructure through which internet messages travel (Figure 10).

Private infrastructure powers the public internet.[17] In the 2000s, debates about net neutrality painted bleak futures in which telecom companies like Comcast would partner with service providers to selectively filter content. "Facebook is free; if you'll want HBO, you'll have to pay us more." In response, the content providers became telecoms. They built private infrastructure, to which no expectation of neutrality applied.[18]

[15] Richard Barbrook and Andy Cameron. The Californian Ideology. 1995.

[16] Paul Dourish. Not the internet, but this internet: how othernets illuminate our feudal internet. In *Proceedings of The Fifth Decennial Aarhus Conference on Critical Alternatives*, pages 157–168, 2015

As Dourish (2015) observes, this internet is neofeudal. Public life happens in private domains, within which there are no market relations.

[17] Todd Arnold et al. Cloud provider connectivity in the flat internet. In *Proceedings of the ACM Internet Measurement Conference*, pages 230–246, 2020

[18] Zachary S Bischof et al. Untangling the world-wide mesh of undersea cables. In *Proceedings of the 17th ACM Workshop on Hot Topics in Networks*, pages 78–84, 2018

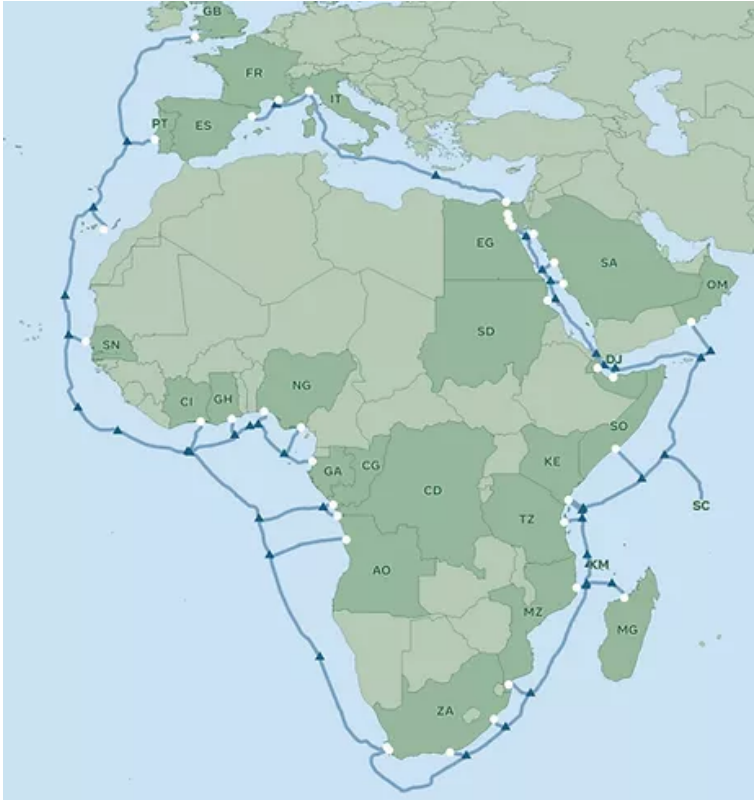

Figure 10: Today, companies like Facebook and Google build their own undersea cables alongside national governments,' connecting countries throughout the global South. Facebook' 2Africa cable, pictured here, is about the same length as the circumference of the earth. From 2Africacable.com.

In much of the South Pacific—countries like Vanuatu, Fiji, Tonga— the only internet connection you can get is to Facebook products, and nothing else. Remember when Facebook's BGP outage brought down all of their products worldwide for a day?[19] In those countries, for most working people, the internet was off. All of it.

The story of the internet, so far, is this. The U.S. military wanted to beat the Soviets on tech, so they built a communications network that turned out to be more useful than anyone imagined. In the 1990s, a small but influential group of techno-utopians planted a liberatory flag in this network. And businesspeople, influenced by the libertarian aspects of these utopians' ideology, built private companies that flourished and consolidated, and flourished and consolidated. Until those companies became the internet itself. These companies helped the U.S. achieve some degree of growth. They helped the U.S. build a global reach for its products. But, somewhere along the way, the U.S.'s deference to private enterprise got the better of it: today, tech companies—the private fiefdoms the American internet minted— compete for power and influence with the state that birthed them.

[19] Brian Krebs. What Happened to Facebook, Instagram, WhatsApp?. October 4, 2021.

The ongoing conflicts between Facebook and the U.S. government over political news, or between Apple and the U.S. government over encryption are, at their heart, questions about their ability to adequately participate in the U.S.'s regime of securitization.

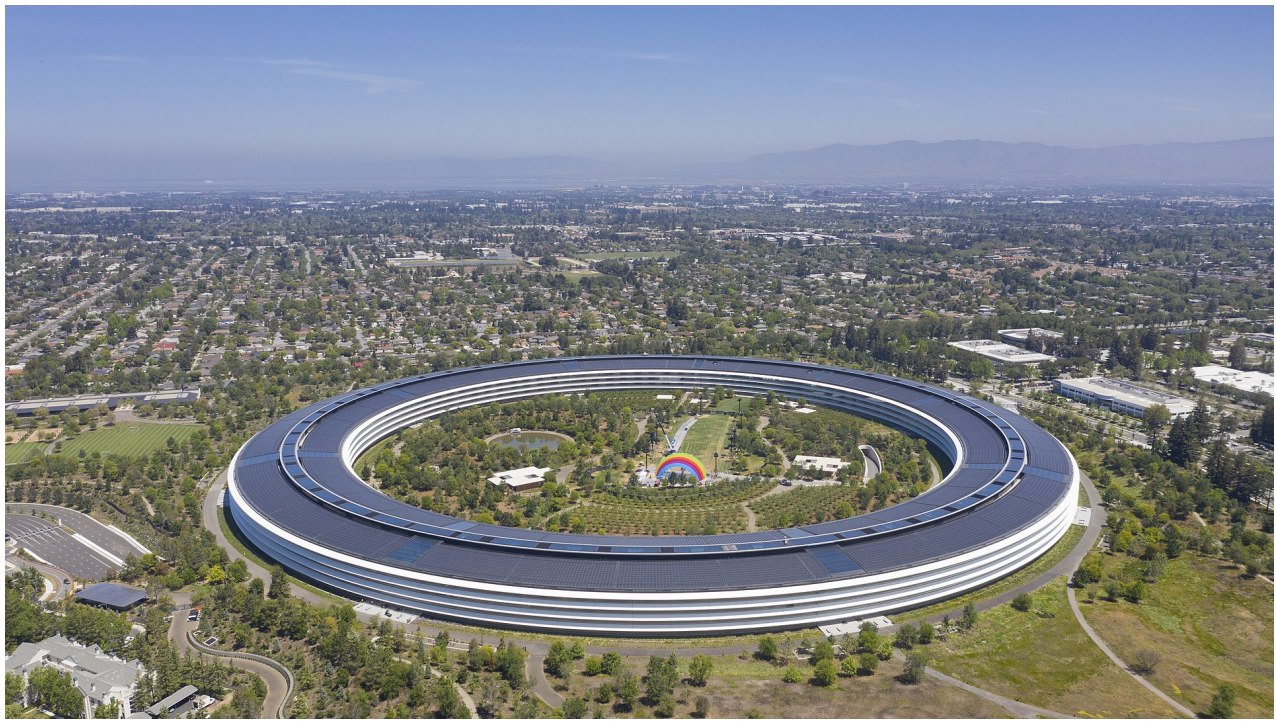

Figure 11: Apple's corporate campus in Cupertino, California.

As John Perry Barlow prophesied, *someone* was granted independence in cyberspace, albeit a partial and complex kind. But it was not us. It was the participants of Davos conferences future, the owners and shareholders of the, mostly, U.S.-domiciled corporations that now provision this internet as we experience it.

## *Why not this internet?*

I've explained why this is the internet we use, why this one became hegemonic. And I've traced the contours of that hegemony in the U.S. public and private sectors.

Now, I'll convince you that this internet, as it exists today, is unsuitable as a global internet. I'll point to two main issues. First, its governance effectively lies in the hands of a handful of U.S.-based corporations and nonprofits, entities that the U.S. can (and does) exert legal power over; this excludes most internet users, including many in the U.S. from participating in meaningful popular oversight regarding the internet they use. Second, non-U.S. national governments' efforts to achieve *their* goals threaten to destabilize and fragment the internet for everyone.

### *Global control points*

Remember: this internet is decentralized, not distributed (Figure 4). That means that, while there is no one center, there are several key *control points*.[20] Points upon which political pressure can be applied.

[20] David D. Clark. Control Point Analysis. *SSRN Electronic Journal*, 2012

Some of that pressure only yields results locally (like blocking websites within your territory, which is what China does with its Great Firewall). But some are global. Some control points affect the entire world. Those control points represent the parts of this internet that are globally shared.

Imagine: instead of blocking a website in China, you block it for everyone. That's a global control point.

What are these control points, and who controls them?

One perspective on this is: what happens when you visit a website? This (Figure 12) is what happens. Don't read this. The point is that it's overwhelming. We have to simplify. We have to identify a finite number of types of institutions that everyone relies on.

Here are six (Figure 13). I identified these institutions earlier this year, working with the Internet Society. Remember: forget your local ISP, like Comcast, who can maybe decide what *you* see. These are institutions that can decide what *everyone* sees.

Now, *of* those institutions, what proportion of them *by marketshare* are based in the United States? Taking the most popular websites in the world, what proportion of them use service providers based in the United States?

The answer, as you can see, is "a lot." In fact, those institutions that it's arguably *easiest* to route around—web hosts, data centers, for which the services are mostly fungible and you can move providers easily enough—those industries are *less* U.S.-dominated. The troubling ones are those it's *hard* to route around, like naming.

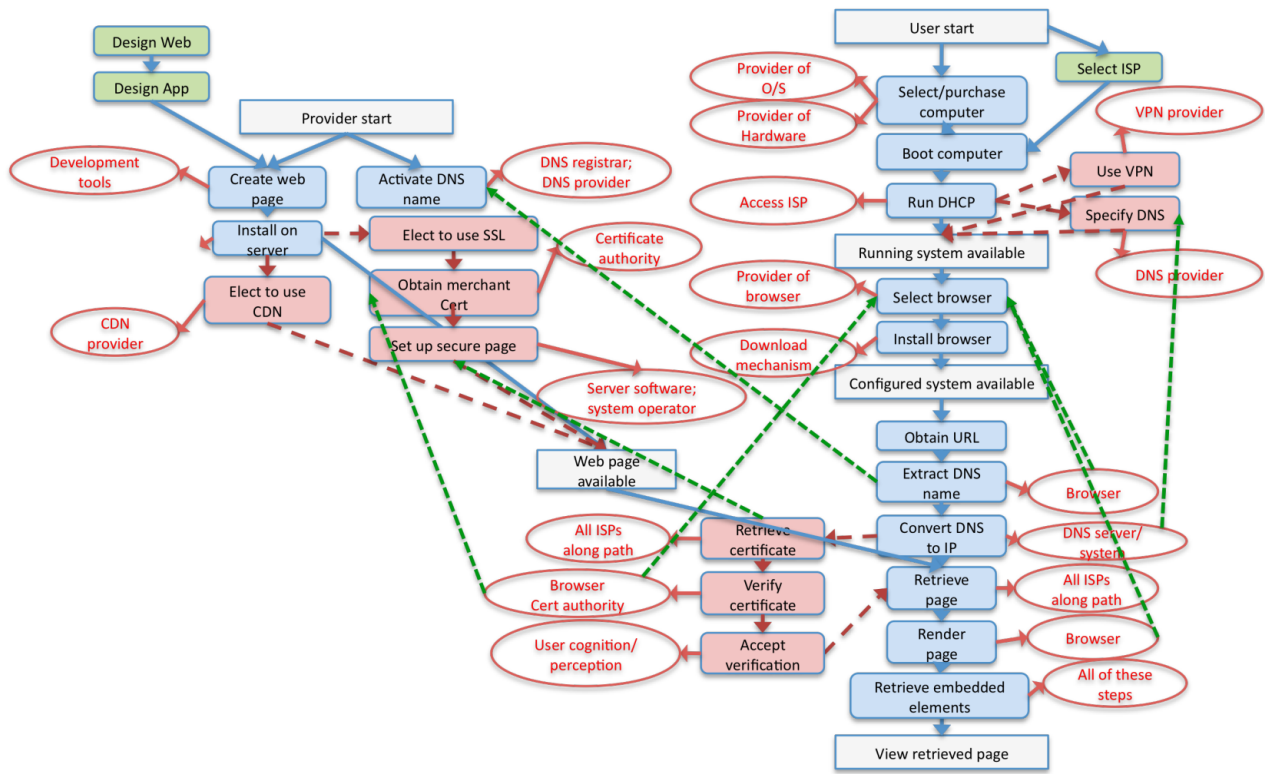

Figure 12: How do users retrieve a webpage? How do designers make one? Blue arrows indicate a sequence of steps. Green arrows capture dependencies between steps. Red ovals are "control points:" actors who can control the outcome of particular steps. Adapted from Clark, 2012.

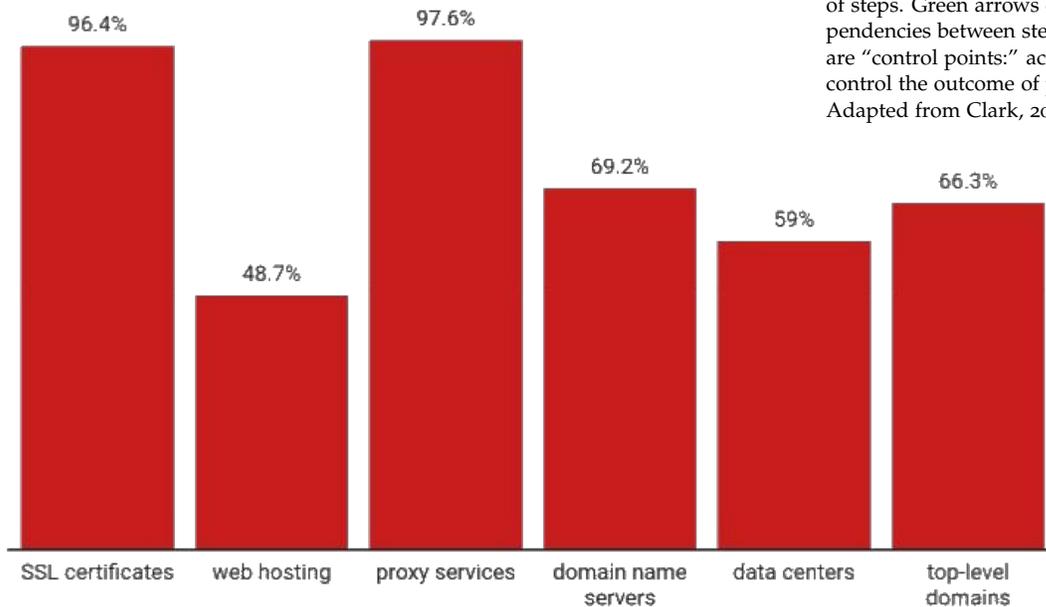

Figure 13: The proportion of core internet services provided by U.S.-based companies by marketshare. The Conversation, Fight for control threatens to destabilize and fragment the internet.

*Domain names*

Let's start with naming.

I love domain names. The benefits of global domain names are huge: *one* human-readable string maps to exactly *one* resource—for the entire world! It's amazing.

`nickmerrill`.`substack`.`com`
*subdomain     domain     TLD*

Figure 14: A typical DNS name, split into its component parts.

How do you get one? Well, first you go to a domain registrar. You go to a company like GoDaddy and say, "I want substack.com, please." So they go to a *domain registry backend* for the ".com" top-level domain (or TLD), and say, "hey, are there any domain names called "substack" in here?" If the registry backend says, "no," the registrar can go on and sell the domain.

But these domain registrars *and* these domain registry backends are all companies or nonprofit organizations, and they're all based somewhere. In fact, by volume, 66.3% of websites *in the world* (and 80% of the world's fifty most popular websites[21]) are registered on top-level domains whose registry backend is domiciled in the United States.

[21] GitHub. daylight-lab/alexa.

So, if you're the U.S. federal government, you can, in theory, issue a *court order* to one of those registrars or registry backends say, "hey, give me this domain name; it's been involved in a crime."

Which is exactly what it does (Figure 15).

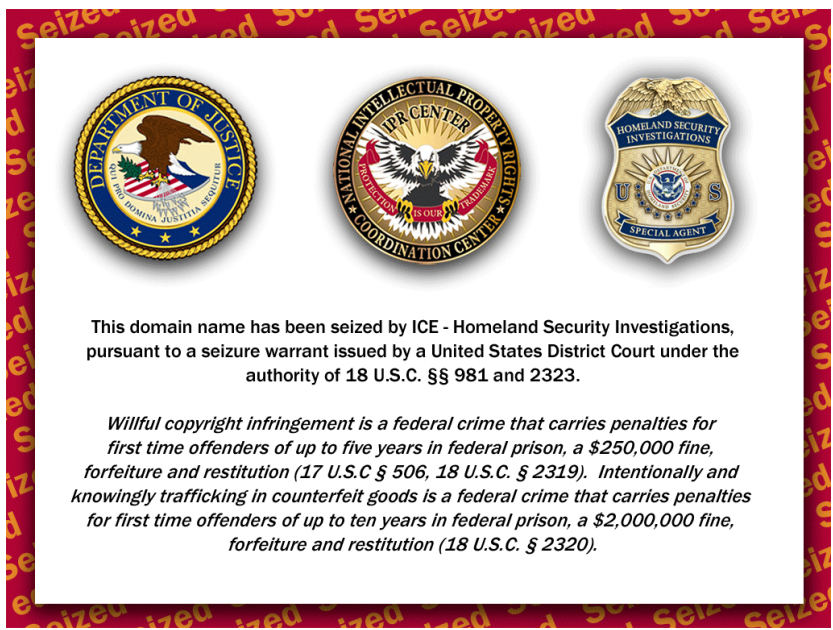

Figure 15: The site ICE redirects you to after seizing your domain. See it for yourself: chinaseatbelt.com

An ICE-administered operation, Operation In Our Sites, has seized 1.2 million websites so far by going to registrars or backends with court orders. This has sometimes meant seizing US citizen's domain names and holding them without due process![22]

Anywhere in the world, you can own a domain name, and that domain name can be seized by the U.S. government if you've registered it with a TLD for which a U.S.-based organization is the registry backend—the case for all of the world's most popular TLDs (.com, .net, .org).

If your website is seized, it still exists, but no one knows how to reach it because they only know it *as* "chinaseatbelt.com" (or whatever your domain was). So it may as well not exist. There's no name by which to find it.

This operation has mostly targeted copyright infringement. But, earlier this year, *Operation In Our Sites* seized an Iranian news site, alleging "disinformation."[23] So the use of this mechanism as an *ideological* tool—indeed, a tool of U.S. interests of all forms, rather than a narrow trade concern—lingers on the horizon.

[22] Karen Kopel. Operation seizing our sites: How the federal government is taking domain names without prior notice. *Berkeley Tech. LJ,* 28:859, 2013

[23] CNN. US government seizes dozens of US website domains connected to Iran. June 23, 2021

### Certificate authorities

So far, the federal government has only seized TLDs, but I have a strong hunch that, at some point, we're going to see this court order technique applied to other sorts of internet infrastructure.

Consider SSL certificates. Ever see that lock in your browser? That lock appears because somewhere, a certificate authority (or CA) issued a valid certificate for that site. That site uses that certificate to encrypt its traffic to you, and your browser looks at the certificate, checking it against the received webpage, and says, "Yep, this looks like it *is* the website, because this authority vouches for them. Here's the decrypted page." It shows you that lock to let you know you're not being phished.

The problem is that the CA system has an extremely questionable design: a CA can revoke any issued certificate at any time. So, the U.S. government could, in theory, go to the CA for a particular website and say, "hey revoke this certificate." And the user would see something like this (Figure 16).

A CA can *also* issue a a valid certificate for any site. Imagine China decided to no longer play by the rules. It could issue a perfectly valid certificate for the New York Times! How do we deal with this problem? By limiting who gets to be a CA. In practice, this has concentrated power among a few, US-based providers. Why? Well, the ultimate decision regarding which CAs to trust lies with web browsers, who ship lists of pre-trusted CAs. And U.S. vendors dominate the web browser market. As a result, the US almost fully controls the global supply of TLS certificates, as you can see!

### Content distribution networks

The real power in today's internet is held by content distribution networks, or CDNs. They include players like Akami, Cloudflare, and Fastly.[24]

CDNs, in broad strokes sit between you, the user, and the content you're requesting. For example, when someone requests spotify.com, their request will in fact resolve to a CDN like Cloudflare, which will to handle the incoming traffic on Spotify's behalf.

Why? Well, incoming requests can sometimes be hostile. Dis-

[24] "Reverse proxies" in Figure 13. Nick Merrill. Cache Rules Everything Around Me (C.R.E.A.M.

I am on the record saying that Cloudflare should be nationalized. The reason I'm hard on them is that they're *so* ripe for both political control *and* cyberattacks.

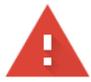

Figure 16: The error your browser will disaply when you visit a site for which the CA has revoked the SSL certificate.

tributed denial of service (DDoS) attacks bombard websites with requests until they can't respond to legitimate requests, effectively making them unavailable. CDNs can block DDoS attacks by observing internet behavior at large scales. Meanwhile, for well-intentioned users, CDNs put the content close to you. CDNs cache frequently accessed content, placing copies in servers around the world. When you request that content, they deliver the version closest to you geographically, minimizing latency.

There's nothing intrinsically wrong with reverse proxies. They provide collective defense against DDoS attacks—using large-scale observation to identify and block decentralized threats. (In the early 2010s, the heyday of Anonymous's DDoS-oriented "hacktivism," some thought DDoS attacks would be an existential threat to the Internet. Services like Cloudflare have largely neutralized that threat.).

The problem with CDNs is first that they are highly centralized: there are only fourteen of them, one of which (Cloudflare) has 80% of the marketshare, and are overwhelmingly (97.6% by marketshare) based in the United States; and second, that they are highly critical: they can effectively censor the Internet.

For example: have you ever used Tor? If you have, you've probably noticed what a pain it is to browse with. You get these CAPTCHA requests every page you visit (Figure 17).

Those CAPTCHAs come from CDNs. Perhaps there's a good reason; perhaps they prevent DDoS attacks.[25] Gating Tor traffic also makes the regular, non-private way of browsing infinitely more convenient, which is *definitely* good for Cloudflare's customers, who

Tor anonymizes traffic: it obfuscates the destination of internet traffic for interlocutors, and its source for recipients. Tor was funded in part by Radio Free Asia to circumvent Chinese censorship.

[25] Read Cloudflare CEO Matthew Prince, The Trouble with Tor.

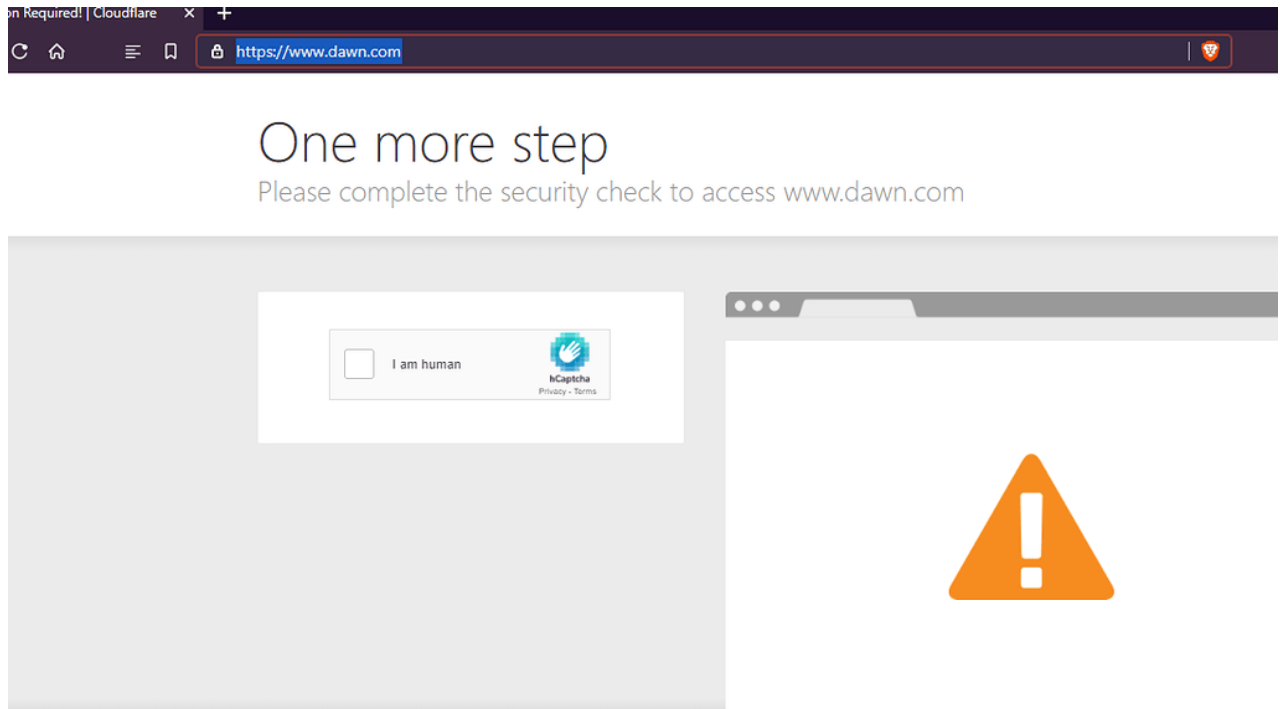

Figure 17: A CAPTCHA you might see while browsing the internet with Tor. CDNs can manipulate traffic in fine-grained ways, targeting the manipulation to specific devices, locations, or people. And they can do so for a tremendous proportion of the internet—particularly the internet's most popular sites. This makes them ripe as a target for political influence and cyberattacks.

overwhelmingly make money from targeted advertising. But my purpose is not to opine on whether Cloudflare is right or wrong in gating Tor traffic. It is to demonstrate that CDNs can largely gate the internet itself, and do so in a fine-grained way, limiting people based on their location, the content they are requesting, and their browsing history elsewhere on the web.

From the perspective of the average user, CDNs *are* the internet. When Fastly went down, a *lot* of the internet went down with it: CNN, Amazon, the UK government's webpage...[26] Now consider: Fastly only has 5% of the CDN market. Cloudflare has 81%. Forget what the U.S. government could accomplish were it to issue court orders to these CDNs. What havoc could a sophisticated, state-backed cyberattack wreak?

[26] See news coverage of outage at the time: CNN. Two obscure service providers briefly broke the internet. It could happen again. June 9th, 2021.

*Tussle*

It is not a question of whether, but how the U.S. federal government will exert legal power over its tech companies to achieve international and domestic aims. What *kind* of power will it wield, and toward what end? Will tech companies accede to its demands, as domain registrars have to Operation In Our Sites?

We can't know. What we *can* know is that global control points exist, and that whoever controls those control points more or less controls what content people can access. Tussle[27] over them will decide what the internet looks like for everyone.

Now. While the internet's global control points lie primarily in the U.S. governments, and companies, and people, worldwide have their complaints about this internet. They don't like what's on it, they don't like what's kept off it. They have domestic and international goals to accomplish, in other words, and the internet is one of many fronts on which they hope to enact them.

What is a government lacking the U.S. direct, jurisdictional control to do? Within their own boundaries, nations' control over domestic physical infrastructure allows them to filter content, throttle it,[28] or just turn the internet off completely, a strategy we've seen most frequently as a means of protest control in India.[29]

From sophisticated censorship regimes like China's to immature ones like Azerbaijan's, the core logic of all internet "sovreignty" is this: you can take what you want from the global internet. In other words, you can interoperate *selectively* with this internet, blocking websites you don't like, and building alternatives internally as you're able. Baidu, for example, is a domestic alternative to Google. Create your own control points domestically.

[27] David D Clark et al. Tussle in cyberspace: defining tomorrow's internet. In *Proceedings of the 2002 conference on Applications, technologies, architectures, and protocols for computer communications*, pages 347–356, 2002

[28] Pengxiong Zhu et al. Characterizing Transnational Internet Performance and the Great Bottleneck of China. *Proceedings of the ACM on Measurement and Analysis of Computing Systems*, 4(1): 1–23, 2020

[29] Rajat Kathuria et al. The Anatomy of An Internet Blackout: Measuring the Economic Impact of Internet Shutdowns in India. *Indian Council for Research on International Economic Relations*, 2018

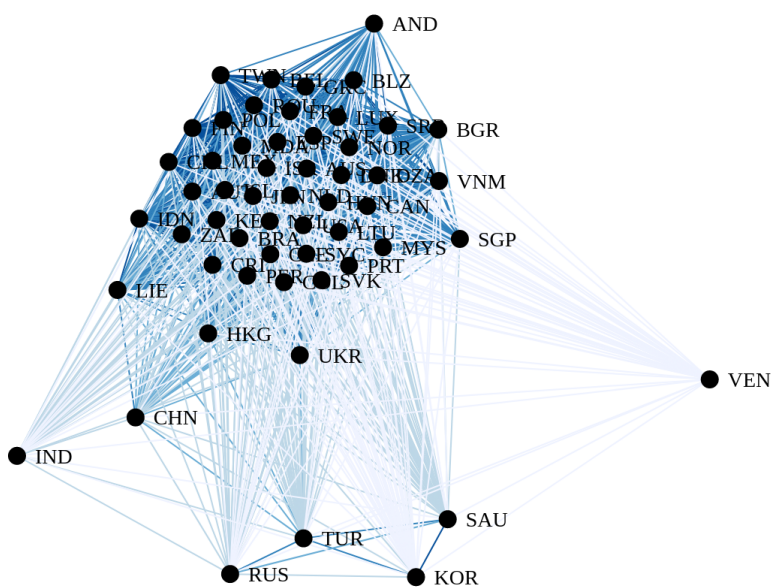

Figure 18: Similarities between countries in the types of web content they block. These similarities are strongly predictive of geopolitical relationships (trade agreements, military alliances, etc). Explore an interactive version of this visualization.

In 2021, Steve Weber and I found that the websites a country blocks reflects its real-world alliances and commitments (Figure 18).[30] We collected data on what websites were blocked everywhere

[30] Nick Merrill and Steven Weber. Web site blocking as a proxy of policy alignment. *First Monday*, 2021

in the world, and compared each country to each other country in the *types* of content they censored (categories like "gambling, "drugs", "politics", etc..). We found that similarities in the websites a country blocks are strongly predictive of who that country trades with, who they are in military alliances with, and so on.

My interpretation of this result is twofold. First, we *already* have a multiplicity of internets, interoperating selectively. (Content blocking *is* selective interop). "Blocs"[31] are emerging (Figure 18).

Second, as Douzet (2014) identified, struggles over the internet both reflect and shape power conflict *broadly*.[32] The internet is, in other words, constitutive of geopolitics. A lever to further it, a driver of it, and a stage on which it plays out. Just like it was back when ARPA invented it.

But today, there are a lot more stakeholders at the table. States, corporations, and other interests compete and cooperate on this stage. And internet sovereignty—controlling the internet within one's own borders—has its limits. States have international goals, as well.

As discussed, the internet is fragile.[33] While targeted attacks are costly to mount, chaos is cheap. Large-scale internet destabilization is one of the clearest routes for a nation to achieve international goals. For example, with a sufficiently resilient domestic internet, a state like Russia[34] may have little to lose from destabilizing rest of the world's internet. There is only one reason, in my mind, why a cyberattack has not yet brought down the whole internet: no one knows how the conflict would escalate. Eventually, someone will fuck around and find out.

What we have today is a dangerous, precarious situation. What comes next?

- *Cyberwar*? Ad-hoc conflict that makes the internet unreliable for everyone? Perhaps ending in a barren, unusable web?

- *Fragmentation*? Siloed internets, fragmented into national "blocs" that trade together, countries outside that bloc only partially or imperfectly reachable?

- Or *hegemony*? An internet that's global, but globally censored and surveilled? Total control, a clamping down by the U.S. state apparatus on this internet. Perhaps riding it out until this internet becomes... someone else's hegemony. China's hegemony?[35] A corporation like Meta's hegemony?[36]

---

[31] Steven Weber. *Bloc by Bloc: How to Build a Global Enterprise for the New Regional Order*. Harvard University Press, 2019

[32] Frédérick Douzet. La géopolitique pour comprendre le cyberespace. *Hérodote*, (1):3–21, 2014; and Kevin Limonier et al. Mapping the Routes of the Internet for Geopolitics: The case of Eastern Ukraine. *First Monday*, 26(5), Apr. 2021

[33] Brian Krebs. The Internet is Held Together With Spit & Baling Wire. November 26, 2021.

[34] Reuters. Russia disconnects from internet in tests as it bolsters security. July 22, 2021.

Louis Pouzin, an early internet luminary, said of this internet, "Les bases que nous avons jetées sont complètement obsolètes." The foundations we have laid are completely obsolete. Les Echos. L'Internet doit être refait de fond en comble. May 2013.

[35] Australian Strategic Policy Institute. China's cyber vision. November 24, 2021.

[36] Or a consortium of corporations, like in the 1992 novel Snow Crash, in which the US has fallen completely to private interests, which have split the U.S. up into privatized fiefdoms? That book coined the term "metaverse." It undoubtedly inspired Mark Zuckerberg to build one. He must have thought, "hey, *we* could be one of those corporations that takes over after the fall of the U.S.!"

*What internets come next?*

LISTEN. *None* of those internets sound good to me.

And *this* internet was never built with popular sovereignty[37] in mind. Powerful institutions built it to do their power better. This internet is serving its goal in the main.

Phrased differently, *the issue isn't that control points exist; it's that there is no popular governance over them.* That you have no say over the control points that affect *you*. While *some* aspects of the internet are heavily democratized in non-governmental bodies,[38] points of global control have mostly been nestled away inside legal entities which are in turn nested inside securitized states. There is no place in these control points for interests that are neither state-required nor profit-motivated. There is no room for local communities. There is no room for local governance. There is no room for popular veto. Except to stop using this internet, which isn't really a choice. How would you get paid?

*Blockchains*

At this point, I have to address blockchains.

Blockchains, like bitcoin, rely *only* on the protocol layer of the internet. They work on top of IP and BGP. While they *do not* replace the internet's most fundamental, material infrastructure—the physical media by which computers communicate over space—they *do* throw this internet's application stack out the window. They provide another possible application stack.

That in and of itself—creating an alternative application layer—is kind of amazing. I hate the way bitcoin does it, with proof of work, an ongoing environmental disaster. I hate that Ethereum, which aims to provide the common good of reproducible computation at scale, seems to want to become another hegemon. The *one chain*. And, of course, I hate blockchain's roaring-20s speculative public face. Eric Adams getting his first three checks as New York's mayor in Bitcoin. The public UC Berkeley funding its operation through NFTs of the CRISPR patent. Decentralized autonomous organizations (DAOs) that try to buy an original U.S. constitution because scroll emoji.[39]

But none of those is my complaint about blockchain here. My complaint about blockchain is that even those developers who are well-meaning, which is some vanishingly small subset of the community, believe a naive technological determinism at times. To paraphrase Amir Taaki, they want to do the revolution, but they think they can party while they do it, because their tech will do the revolu-

---

[37] Milton Mueller. *Will the internet fragment?: Sovereignty, globalization and cyberspace*. John Wiley & Sons, 2017

It got away from its original designers in the U.S. military in some ways, that are surprising but, perhaps not *too* surprising: the U.S. military apparatus, the US higher-ed system, and the U.S. private industrial apparatus are *all* quite cozy with each other.

[38] Corinne Cath. The technology we choose to create: Human rights advocacy in the Internet Engineering Task Force. *Telecommunications Policy*, 45 (6):102144, 2021

I'm sorry. I wanted not to.

@ericadamsfornyc on Twitter; UC Berkeley news; Bloomberg.

[39] Ali Breiland at Mother Jones described this speculative phenomenon as the only rational response to an irrational market: these NFTs may be a scam, but so is the economy.

tion for them.[40]

That's never how it works. That did not work for the techno utopians. And if this community, even the *good* people in this community, aren't careful, they will deliver their technologies to the same state and capital interests as their predecessors.

[40] The Blockchain Socialist. Let there be DarkFi and Anonymity.

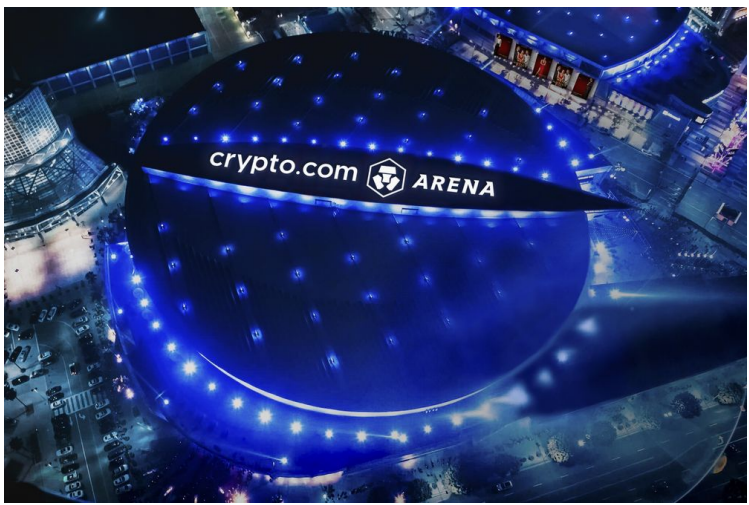

Figure 19: The Crypto.com Arena.

*Manynets*

As Buchmann (2018) remarked, the original network of networks was the empire.[41] Empire connected smaller societies and forced them to interoperate. That was its innovation. You can extract surplus value that way! You can make a hegemon.

[41] Ethan Buchmann. A brief history of distributed state. July 12, 2018.

In many ways, this internet is that empire taken to its logical conclusion. It connects, with caveats, all of the world's networks into a single network of networks.

At one point in the past, perhaps this singular internet served the world better than it does today. *Maybe*. But, today this, internet doesn't please anyone, let alone everyone. It doesn't respond to the needs of people on the ground. Empires that stop responding to the needs of people on the ground they have a tendency to break apart.

Here's a poll I did on twitter (Figure 20). Unscientific poll. But it seems the experts agree: we're heading toward—in some way, in some sense—a multiplicity of internets. The only question is, whose internets will they be?

The states'?
The corporations'?
Or the peoples'?

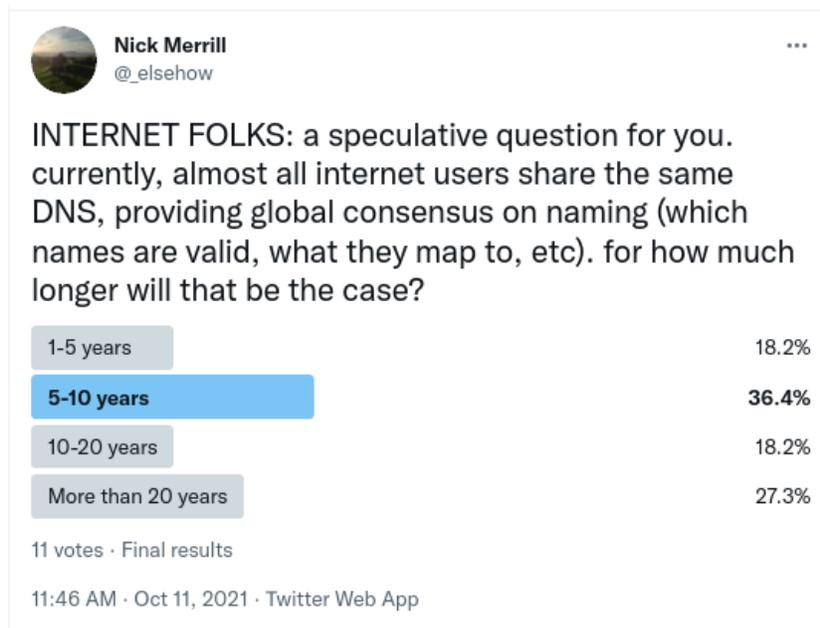

Figure 20: My Twitter followers (broadly, internet governance folks and academics) are pessimistic that our global system for assigning mutable names will remain global for long. 54% think it won't last beyond another ten years. Only a minority think it will survive more than twenty.

*A multiplicity of internets*

There is no rule that there can only be *one* internet.

In fact, if there's any one thing that binds together almost all of the world's internet-using population under one experience, it's that no matter where you are, there is only *one* internet on offer. You don't get to choose which internet you see. Whatever choice of service providers you have, there will be some finite amount of content available to you. Unless you're really rich, or really crafty, that's the internet you're getting.

No matter who you talk to in any country, no matter how different their internet is to yours, everyone calls it "the" internet. Like "the" subway. In the world, there may be many, but to you, there's only one.

This, is in some sense, the purest form of control. The control of no alternative.

What would alternatives look like?

They might look like this (Figure 21). NYC Mesh, pictured here, seeks to provide affordable, sliding-scale internet service, faster and at a higher-quality than the local telecom monopoly.

Others, like the Mycelium Mesh Network, built in the wake of the BLM protests of 2020, provides a backup network. Resilience, in case law enforcement cuts telecommunications infrastructure to control a protest.[42]

Remember: this internet may be global, but *your experience of it* is always local. [43] The internet *to you* is equivalent to *its final hop* to you. From your perspective, this internet is *always* on the ground. That basic fact can make this internet different in different places, particularly in authoritarian countries with tight controls on speech. But it also opens up opportunities to build *new* internets. What to you just looks like a normal WiFi network may in fact connect you to a completely different technological stack.

[42] A strategy pioneered by the San Francisco BART police in 2012. Vice. Activists are Designing Mesh Networks to Deploy During Civil Unrest. October 5, 2021.
[43] Janet Abbate. What and where is the Internet?(Re) defining Internet histories. *Internet Histories*, 1(1-2):8–14, 2017

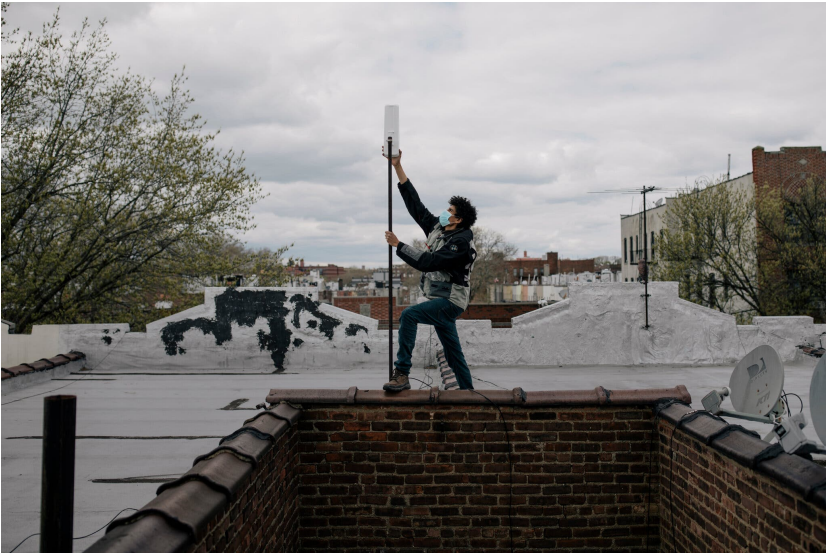

Figure 21: NYC Mesh, New York City's community-run internet service provider. A small set of volunteers build radio towers on rooftops, manage routing tables, for the many who enjoy the service. Image: The New York Times. 'Welcome to the Mesh, Brother': Guerrilla Wi-Fi Comes to New York. July 16, 2021.

The limit is our ability to organize.

At some scale of organization, there's this (Figure 22). Guifi.net, a mesh network centered in Catalonia. A peer-to-peer ISP, explicitly left in its ideology, that blends community-run internet with goals for an independent Catalonia. Guifi.net connects about 37 thousand machines. It has no CEOs, no leaders. Anyone can build infrastructure and extend the network. The infrastructure they build becomes part of the commons.

*The commons*

Speaking of the commons. Mueller (2017) argues that the internet will *never* fragment because a global, shared communications network is too valuable to *everyone*. [44] Like the climate, it's a good we can't do without.

I don't disagree! But what parts of that network do we all need to agree on? All of it? *Everything*?

Or is it possible to produce something more confederated? A network of networks that connects globally, but voluntarily. A global commons that's locally governed.

That's what I'm interested in now. How to build *new* internets we can live with, fall back on, interoperate between. Both technically and on the ground—how to organize to make robust communities.

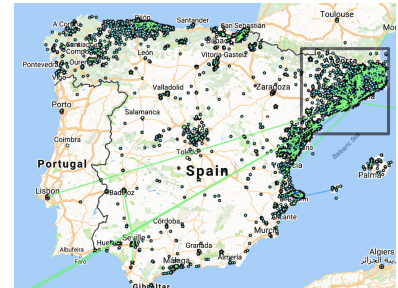

Figure 22: Guifi.net, a peer-to-peer ISP that connects about 37 thousand machines.

[44] Milton Mueller. *Will the internet fragment?: Sovereignty, globalization and cyberspace*. John Wiley & Sons, 2017

It was a long, circuitous route to this topic for me. I lost a lot of sleep over how to protect this internet. Now, I'm not so sure this internet is worth saving. Or, rather, I don't think that mission *as such* is the most tenable or even the most effective goal.

The better goal is to make internets that work better at much smaller scales. At urban scales, the scale of towns and cities, what Murray Bookchin calls the "forefront of political life." [45]

I can envision a multiplicity of internets stitched together, a global confederalism. They can be local, self-sovereign, managed popularly within communities. They can interoperate carefully, pragmatically, with this internet, and, eventually, with one another.

I'm not saying every place in the world will have such an internet. I'm saying that some place *could*.

And we can invent infrastructure to power those internets. We can invent institutions to govern those infrastructures. But it's not just about tech. It's about the organizing. It's about the story. We use this internet because it cohered to stories that were *legible as liberation at the time*. Yes, that flavor of liberation was of a distinctly neoliberal, late-20th-century variety. It was a product of its time. But that doesn't have to be the end of history, the last internet, the one internet to rule them all forever. In fact, it won't be.

The problem is that no better internet has come next, yet.

From paranets for protesters to StreetNets in Havana, it seems to me that an *ideologically* trustworthy communications network will be a necessary pre-requisite for any meaningfully liberatory struggle we might hope to achieve in our lifetimes. That network will be an internet. But perhaps not this internet.

---

I went through a long phase of trying to figure out where this internet was vulnerable, what kinds of outages could cascade into catastrophic failures that would cause global trade to desynchronize.

[45] M. Bookchin. *Urbanization Without Cities: The Rise and Decline of Citizenship*. Black Rose Books, No. V171. Black Rose Books, 1992. ISBN 9781895431018

## Q&A

**Colleague:** I buy everything that you said so far about corporate and state control over the existing internet, about the need for alternatives, but I'm trying to square the challenges of cybersecurity with our commitment to building local internets. Don't we still have huge security challenges that may be beyond the scale of, for example, a community in Catalonia?

**Nick:** I certainly can't pretend that making local internets will improve security unilaterally, or by default. But there are two things I'd like to point out here.

First, this internet, by design, is not terribly secure. I've mentioned that routing issue earlier. We have issues with the most fundamental layers of the internet stack, and with BGP in particular.[46] That's a structural issue that isn't going away. So, in some sense, building new internets isn't the craziest thing from a cybersecurity perspective. We *do* need some of these things to go away.

Second, on the risk that new internets will be insecure: of course, they *will* have issues, they *will* be vulnerable. But one of the biggest, affordances for attackers today is that this internet is homogeneous. It is the same everywhere. If you find an exploit, it will affect everyone.

When Baran was talking back in the day about decentralized and distributed networks (Figure 4) he was thinking physically: you bomb something. What emerged instead is a network that is decentralized physically, but logically totally centralized. That is, in fact, the core of a lot of cybersecurity issues.[47]

So there are security trade-offs here. The received wisdom is that fragmenting the internet will make it less secure. I don't think that's all of the story. In fact, I think worrying about those trade-offs is a bit academic in an internet that is insecure by design. That can be taken down by a sophisticated cyber attack, or disabled because the U.S. government feels like it.

**Student:** Who would build these decentralized internets? If there were to be a catalyst, would you see this at a local city level, like Berkeley, or at a university level? How could this start mobilizing, in your mind, such that it's sustainable and financed and maintained and all these sorts of things?

**Nick:** I think what you're asking is, in many ways, the million dollar question.

Yes, I think that the municipal level is the place to begin. You have people in Berkeley, like Ben Bartlett, who are interested in blockchain.[48] Ben Bartlett in particular is interested in doing com-munity bonds and the like. I think there's energy among people like him, throughout the country and world, that can be reoriented from blockchain narrowly to something like this.

---

[46] When BGP was invented, we didn't have authenticated ways for announcing routes or finding Byzantine fault tolerant consensus. Now, we're stuck with BGP, because we'd have to upgrade everything at the same time globally or the internet would globally break!

[47] We observe this effect from core internet infrastructure all the way up to the application layer. For example, bugs in Microsoft Exchange Server: Microsoft Exchange Server is self-hosted, so it's decentralized, but a bug in the implementation affects all the decentralized components. (NPR, China's Microsoft Hack May Have Had A Bigger Purpose Than Just Spying. August 26, 2021.).

[48] Ben Bartlett's efforts were written up in Wired. How a Blockchain Could Help Roll Out Berkeley's Next Fire Truck. July 9, 2019.

These projects hook into other debates that may be legible to those policy makers: things like municipal WiFi, municipal broadband. Local leaders care about those things, there's energy around them. And I think when you have energy, and you have access to telephone poles, a lot can happen.

The question remains about how to organize on the ground, though. How to catalyze municipal action. Those questions aren't going to be resolved by us. That's only going to be resolved in the usual way, the messy way, the democratic way. But I look forward in partaking in those discussions. That's the real work, in my opinion.

**Student:** Oftentimes, organizing on the ground is most effective when there's money behind it. There are think tanks and organizations behind certain grassroots initiatives. Sometimes, when it comes to these sort of topics and the issues that would likely benefit society as a whole over corporations, money is missing. Where would the money for something like this come from?

**Nick:** There's a great project called Helium. Helium provides cheap, low-power wireless internet connectivity for IoT devices, and it has coverage all throughout the world.[49] Most large cities in North America and Europe have excellent coverage. How did it get community members to coordinate at such a scale?

[49] Helium coverage map. See the coverage in your neighborhood.

It used a clever mechanism where, instead of proof of work or proof of stake, it relies on *proof of coverage*: you prove that you cover a certain area with wireless connectivity, and you get rewarded in the native token of the Helium network. That token is fungible, you can trade it with other tokens, and that brings liquidity to the people who power this work.

These tokens can organize people because the tokens can be traded for U.S. dollars.

I do believe, with caveats, in this style of tokenized system to deliver those incentives. The proof, in my mind, is in the pudding. Look at Helium and see their coverage. This is not just like some number on a screen. This is the effect of coordination *around* some number on a screen.

As much as I was, for years, a huge skeptic about "cryptocurrencies" (which I now call tokens), I have come around to their power to coordinate collective action. And projects like Cosmos, things in that ecosystem, provide tools for democratic decision-making around how to allocate those incentives. That's the true power of decentralized autonomous organizations, or DAOs: they provide financial incentives, but also provide mechanisms for popular governance over how those incentives are allocated, and to whom.

**Student:** A problem with confederated internets built around communities is that they might reinforce caste-, race- and religion-based silos. One of the checks against these traditional, regressive trends is federation—a democratic state—-particularly in countries like India. How does this decentralized model protect us against this?

**Nick:** Can this new model go on to reify the same human-against-human oppressive systems they oppose? To me, this is the key question in all of politics. How do we make progress against a political system that subjugates human under human? As Bookchin observes, our subjugation, our exploitation of the environment stem naturally from our desire to subjugate humans. We subjugate nature because we're so comfortable subjugating humans. This is the start of it all—our being okay with subjugation. So how do we make progress against this?

I don't know the answer to this, but clearly the answer is on the ground somewhere. And your question highlights how local these issues are, how situated they are. These technologies never come without ideological baggage. The key question, I think, is how do we attach the right kinds of ideological baggage in rolling out new internets? How do we make it harder to subjugate? In this context, this caste, race, religion context in India, I would love to think more about this with you.

**Student:** Who is able to hide more easily within distributed internets? For example, to protect data that is not covered by end-to-end encryption, the chat app Telegram uses a distributed infrastructure: chat data is stored in multiple data centers around the globe, controlled by different legal entities, spread across different jurisdictions. The relevant decryption keys are split into parts and never kept in the same place as the data they protect. As a result, several court orders from different jurisdictions are required to force us to give up any data. No one single government can access it.

But, as we understand, white nationalist groups and other extremist groups thrive on telegram *because* of its distributed model. No real, no government or political entity is able to access this data in a way that secures marginalized groups from attacks coordinated on a distributed internet. What are your concerns about a decentralized, distributed structure?

**Nick:** Your question is a question about securitization. My question is a question about who gets to do securitization. No one among us gets to decide what securitization looks like for *any* piece of infrastructure. Can decentralized systems produce forms of hiding in a way that trouble securitization as defined popularly? Of course. And at least there's some popular governance to correct that. Here, centralized or decentralized systems without oversight allow actors to hide in an ad hoc and chaotic way, in a securitization regime that none of us get to decide anything about. That's the issue.

In this example, the one party that can always access that data is Telegram. They have all the data. They are the one party that can always hide. This is what I mean by corporations competing with states for power.

**Student:** Another avenue is to embed the democratic element inside the company. This has the advantage that the institution already exists. And there are alternate corporate structures, like in Germany, where workers are present at the board of directors, or the works councils there. So maybe this is just about upgrading some of the legal code and corporate code to include this democratic element inside the corporations itself. Is that aligned with the direction you want to see?

**Nick:** To take companies that already exist and somehow subvert their governance structure to make them democratic: how? What's the path to that outcome?

The path to a localized internet is clearer to me: I know, or I'm highly confident, that people are only going to become more dissatisfied with the internet that they see. They can change that fact on the ground, at a local level, and inter-operate with other internets as they please.

Can we, perhaps simultaneously, reform existing institutions? Of course. But how? What is the work of that? Does it mean building competing institutions that achieve the same function as legacy ones, yet are global and democratic? Or does it mean storming Cloudflare? If the latter, who does the storming? If a popular movement, that could be bloody, chaotic, and not necessarily democratic on the other side. If it's the FBI, or some successor to the FBI under some different regime, how do we build that FBI with the institutions we have?

The local revolution is simple. It's quiet. It's not violent. It happens voluntarily. And, as people like the internet that is local to them, the other internet fades away. They prefer the one that they like, and they use the one that they like, and they use the other one less. That's the power of this localist perspective, I think. It's voluntary all the way down.

*Some questions for reflection*

- "In the coming 10 years, I believe…"

    – The internet will facilitate more harm than good.

    – The internet will facilitate more good than harm.

    – Don't know/Not sure.

- What do grassroots communications look like in a world of regulated radio frequencies?

- How will quantum communication enable or confound the construction of local internet infrastructure?

- Do any logical internet functions *need* to be globally shared globally (i.e., by all internets)? Interface addresses? Mutable names? Content addresses?

*Acknowledgements*

Thanks to AnnaLee Saxenian for prompting me to put these ideas together, and to the students of INFO 291 (Fall 2021) for their incisive questions.

THIS INTERNET, ON THE GROUND 29Francis Fukuyama. *The end of history and the last man*. Simon and Schuster, 1992.

Tung-Hui Hu. Truckstops on the Information Superhighway: Ant Farm, SRI, and the Cloud. *Journal of the New Media Caucus*, 2014.

Rajat Kathuria et al. The Anatomy of An Internet Blackout: Measuring the Economic Impact of Internet Shutdowns in India. *Indian Council for Research on International Economic Relations*, 2018.

Karen Kopel. Operation seizing our sites: How the federal government is taking domain names without prior notice. *Berkeley Tech. LJ*, 28:859, 2013.

Kevin Limonier et al. Mapping the Routes of the Internet for Geopolitics: The case of Eastern Ukraine. *First Monday*, 26(5), Apr. 2021.

Nick Merrill and Steven Weber. Web site blocking as a proxy of policy alignment. *First Monday*, 2021.

Milton Mueller. *Will the internet fragment?: Sovereignty, globalization and cyberspace*. John Wiley & Sons, 2017.

Fred Turner. *From counterculture to cyberculture*. University of Chicago Press, 2010.

Steven Weber. *Bloc by Bloc: How to Build a Global Enterprise for the New Regional Order*. Harvard University Press, 2019.

Pengxiong Zhu et al. Characterizing Transnational Internet Performance and the Great Bottleneck of China. *Proceedings of the ACM on Measurement and Analysis of Computing Systems*, 4(1):1–23, 2020.